\documentclass[aps,preprint,showpacs]{revtex4}

\usepackage{amsmath}
\usepackage{amsfonts}
\usepackage{amssymb}
\usepackage{graphicx}

\begin{document}

\title{\textbf{Rotating Accelerator-Mode Islands}}
\author{\textbf{Oded Barash and Itzhack Dana}}
\affiliation{Minerva Center and Department of Physics, Bar-Ilan University,
Ramat-Gan 52900, Israel}

\begin{abstract}
The existence of rotating accelerator-mode islands (RAIs), performing
quasiregular motion in rotational resonances of order $m>1$ of the standard
map, is firmly established by an accurate numerical analysis of all the
known data. It is found that many accelerator-mode islands for relatively
small nonintegrability parameter $K$ are RAIs visiting resonances of
different orders $m\leq 3$. For sufficiently large $K$, one finds also
``pure'' RAIs visiting only resonances of the {\em same} order, $m=2$ or 
$m=3$. RAIs, even quite small ones, are shown to exhibit sufficient
stickiness to produce an anomalous chaotic transport. The RAIs are basically
different in nature from accelerator-mode islands in resonances of the
``forced'' standard map which was extensively studied recently in the
context of quantum accelerator modes.\newline
\end{abstract}

\pacs{05.45.Ac, 45.05.+x, 05.45.Mt}
\maketitle

\begin{center}
{\bf I. INTRODUCTION}\\[0pt]
\end{center}

During the last three decades, the classical concept of ``accelerator mode''
(AM) has become of central importance in the theory of chaotic transport in
Hamiltonian systems \cite
{bvc,cm,krw,ikh,d,km,dh,ihkm,zen,bkwz,wbkz,ze,sd,shi,d1,kz,bd}. Recently,
this concept has also provided an illuminating explanation \cite
{fgr,qam,qam1} for a purely quantum acceleration of kicked atoms falling
under gravity, observed in atom-optics experiments \cite{ao}. AMs are
generalized periodic orbits (POs) of Hamiltonian maps having a translational
symmetry in phase space. A paradigmatic example is the standard map \cite
{bvc} 
\begin{equation}
\ M:\ \ p_{t+1}=p_{t}+K\sin (x_{t}),\ \ \ \ x_{t+1}=x_{t}+p_{t+1}\ \text{mod}
(2\pi ),  \label{map}
\end{equation}
where $p$ is angular momentum, $x$ is angle, $K$ is a nonintegrability
parameter, and $t$ is the ``integer'' time. The map (\ref{map}) and its
orbit structure are translationally invariant in $p$ with period $2\pi $ on
the cylindrical phase space $-\infty <p<\infty $, $0\leq x<2\pi $. This
allows to define consistently a PO of (minimal) period $n$ of (\ref{map}) in
a generalized fashion: 
\begin{equation}
p_{t+n}=p_{t}+2\pi w,\ \ \ \ x_{t+n}=x_{t},\ \ \   \label{am}
\end{equation}
where $w$ is an integer, the ``jumping index''. For $w=0$, (\ref{am})
corresponds to a usual (closed) PO on the cylinder while for $w\neq 0$ the
PO is an AM with average acceleration $2\pi w/n$ per map iteration. If the
AM is stable, its $n$ points are surrounded by islands. Stickiness to the
boundaries of AM islands can lead to a superdiffusion of the chaotic motion, 
$\left\langle p_{t}^{2}\right\rangle \propto t^{\mu }$, $1<\mu <2$ \cite
{ihkm,zen,bkwz,wbkz,ze,sd}, where $\left\langle \ \right\rangle $ denotes
ensemble average in the chaotic region.\newline

AMs can arise only for sufficiently large $K$, $K>K_{{\rm c}}\approx 0.9716$, 
when no rotational tori exist \cite{jg} and unbounded motion in the $p$
direction becomes then possible. Thus, AM islands are basically different in
nature from the well-known rotational-resonance islands which exist for
arbitrarily small $K$. The latter islands form the most basic component of
ordered and stable motion in the twist map (\ref{map}). They are associated
with the closed ($w=0$) ``Poincar\'{e}-Birkhoff'' or {\em ``ordered''} POs 
\cite{ll,jdm} which are dynamically equivalent to pure rotations, i.e., the $
K=0$ POs, and emerge from them as $K$ is ``switched on'' \cite{ll}. Despite
the difference above, however, one may expect from the following general
arguments the existence of an interesting kind of AM islands, resembling
rotational-resonance islands in some aspects.\newline

It is known \cite{mmp} that rotational resonances for the standard map (and
similar maps \cite{saw,bd1}) can be constructed in a well-defined way for 
{\em all} $K$; a resonance of order $m$ is a chain of $m$ ``zones'' built on
a hyperbolic ordered PO of period $m$\ (see more details in Sec. II). Strong
numerical evidence \cite{mmp} and exact results \cite{saw,qc} indicate that
for $K>K_{{\rm c}}$ the resonances so constructed give a {\em partition} of
phase space. This implies that an arbitrary orbit of (\ref{map}) consists of 
{\em quasiregular} segments within resonances, where each segment is a piece
of the orbit performing a number of rotations in one resonance \cite{d2,d3}.
Now, a general stability island must lie entirely in some resonance zone 
\cite{bd} and will thus perform a similar quasiregular motion within
resonances. Clearly, the rotational quasiregularity is evident {\em only}
when the island visits resonances of order $m>1$. AM islands visiting $m>1$
resonances are most interesting objects since they exhibit a ``hybrid''
nature: In some time intervals, they rotate like $m>1$ resonance islands in
a near-integrable regime ($K\ll 1$) and at other times they accelerate due
to particular transitions between resonances occurring only for $K>K_{{\rm c}
}$. We thus call these islands, if they exist, ``rotating accelerator-mode
islands'' (RAIs). The RAIs should have a distinct impact on Hamiltonian
transport by generating a new kind of chaotic flight, featuring a
quasiregular steplike structure due to the ``horizontal'' rotation within
resonances. General ideas in Ref. \cite{bd} were illustrated only for the
most well-known AM islands of the standard map, those with central period $
n=1$ which emerge for $K>2\pi $ \cite{bvc,krw,ikh}. These islands lie within 
$m=1$ resonances \cite{bd}. The question of the actual existence of RAIs was
not addressed in Ref. \cite{bd}.\newline

In this paper, the existence of RAIs in the standard map is firmly
established by an accurate numerical analysis, examining also all the known
data on AM islands of which we are aware. In this analysis, the sequence of
resonances visited by an orbit is determined by using the efficient method
introduced in work \cite{d3}. A large fraction of the AM islands for $K<2\pi 
$, listed in Ref. \cite{ikh}, are found to be RAIs visiting resonances of
different orders $m\leq 3$. Some of the significant peaks in the
chaotic-diffusion coefficient observed in Ref. \cite{ikh} for $K<2\pi $ are
due to RAIs. Among all the period-2 AM islands for $2\pi <K<20$, listed in
Ref. \cite{krw}, we have found RAIs visiting resonances of the {\em same}
order $m=2$ (``pure'' $m=2$ RAIs). We discover at $K\approx 8.916$ an
apparently new AM island, a pure $m=3$ RAI. It is shown that even quite
small RAIs exhibit sufficient stickiness to produce an anomalous chaotic
transport. Due to limitations in our available computational resources, we
were not able to find RAIs visiting resonances of order $m>3$. The paper is
organized as follows. In Sec. II, we briefly summarize the notion of
rotational quasiregularity within resonances. In Sec. III, the existence of
RAIs is established by an accurate determination of the quasiregularity
characteristics of many AM islands. In Sec. IV, we briefly study some of the
effects of RAIs on Hamiltonian chaotic transport. A discussion and
conclusions are presented in Sec. V, where we also consider the basic
difference between RAIs and AM\ islands visiting resonances of the
``forced'' standard map [map (\ref{map}) with the addition of a constant
force], which has attracted much attention recently in the context of
``quantum AMs'' \cite{fgr,qam,qam1,ao}.\newline

\begin{center}
{\bf II. ROTATIONAL RESONANCES AND QUASIREGULARITY}\\[0pt]
\end{center}

We briefly summarize here the definition of rotational resonances for the
standard map \cite{mmp} and the notion of quasiregularity within these
resonances \cite{bd,d2,d3}. Let us first recall the concept of rotationally 
{\em ordered} POs \cite{jdm}. In the pure-rotation case of $K=0$, with
constant $p_{t}=p_{0}$, the sequence of orbit angles $x_{t}$ is given by $
x_{t}=x_{0}+p_{0}t$ mod($2\pi $). For rational winding number $\nu
=p_{0}/(2\pi )=l/m$, where $(l,m)$ are coprime integers, the orbit must be a
PO with period $m$; the $m$ PO points are uniformly distributed on the
circle $[0,\ 2\pi )$, i.e., the ``gap'' ${\bf G}_{t}$ between $x_{t}$ and a
neighboring point has the constant width $2\pi /m$, independent of $t$. The
rotational motion on the circle is expressed by the fact that ${\bf G}_{t+1}$
is also a gap, always separated from ${\bf G}_{t}$ by $|l|-1$ gaps. Now, for 
$K\neq 0$, the winding number $\nu $ for a closed ($w=0$) PO is the average
value of $p_{t}/(2\pi )$ and $\nu $ is again rational. A gap is a pair of PO
points having neighboring values of $x_{t}$ in the circle. Then, a
rotationally ordered PO is a closed PO having the two main characteristics
of a $K=0$ PO: (a) $\nu =l/m$, where $m$ is the PO period and $(l,m)$ are
coprime. (b) If ${\bf G}_{t}$ is a gap, ${\bf G}_{t+1}$ is also a gap,
always separated from ${\bf G}_{t}$ by $|l|-1$ gaps. Unlike the case of $K=0$, 
however, the gap width generally depends on $t$.
\newline

An ordered hyperbolic PO with arbitrary winding number $\nu =l/m$ exists for
all $K$ \cite{jdm}. One gap of this PO, say ${\bf G}_{0}$, appears to be
always symmetrically positioned around the ``dominant'' symmetry line $x=\pi 
$, i.e., $\pi -x_{L}=x_{R}-\pi $, where $L$ and $R$ denote the left and
right point, respectively, of ${\bf G}_{0}$. The $l/m$ resonance is now
defined, briefly, as follows (see more details in Refs. \cite{bd,mmp} and
refer to the examples in Fig. 1). One constructs in ${\bf G}_{0}$ a closed
region ${\cal Z}^{(0)}(l/m)$ bounded by four curved segments, which are
suitably chosen pieces of the stable and unstable manifolds of $L$ and $R$
under the map $M^{m}$. The $l/m$ resonance is then the chain of $m$ zones $
{\cal Z}^{(t)}(l/m)=M^{-t}{\cal Z}^{(0)}(l/m)$, $t=0,...,m-1$; see, e.g.,
resonances $0/1$ and $1/2$ in Fig. 1. Clearly, the zone ${\cal Z}
^{(m)}(l/m)=M^{-m}{\cal Z}^{(0)}(l/m)$ lies again in ${\bf G}_{0}$ and
differs from the ``principal'' zone ${\cal Z}^{(0)}(l/m)$ by two {\em 
turnstiles} created by homoclinic oscillations under $M^{-m}$ (the dashed
lines in Fig. 1). Each turnstile consists of two lobes of equal area. By
construction, the lobes outside (inside) ${\cal Z}^{(0)}(l/m)$ form the
region entering (exiting) resonance $l/m$ in one iteration of $M$.\newline

Strong numerical evidence \cite{mmp,d3} and exact results \cite{saw,qc}
indicate that for $K>K_{{\rm c}}\approx 0.9716$ the resonances constructed
as above give, for all $l/m$, a complete partition of phase space. This
implies that a generic orbit must have all its points within resonances and
must therefore perform a quasiregular motion as follows. An initial orbit
point in, say, resonance $l/m$ will ``rotate'', jumping from zone ${\cal Z}
^{(t)}(l/m)$ in gap ${\bf G}_{t}$ to zone ${\cal Z}^{(t+1)}(l/m)$ in gap $
{\bf G}_{t+1}$, until it will arrive to ${\cal Z}^{(0)}(l/m)$. If it does
not lie in an exiting turnstile lobe, it will rotate again, returning to $
{\cal Z}^{(0)}(l/m)$ after $m$ iterations. If, on the other hand, it lies in
an exiting turnstile lobe, more precisely in the {\em overlap} of this lobe
with an entering turnstile lobe of resonance $l^{\prime }/m^{\prime }$ (such
overlaps are the shaded regions in Fig. 1), it will escape to zone ${\cal Z}
^{(m^{\prime }-1)}(l^{\prime }/m^{\prime })$ of $l^{\prime }/m^{\prime }$;
it will then perform at least a finite number of rotations (of $m^{\prime }$
iterations each) in $l^{\prime }/m^{\prime }$ before escaping to another
resonance. Thus, the orbit is a sequence of quasiregular segments, each
lying in some resonance $l_{r}/m_{r}$, $-\infty <r<\infty $, and having a
length of $q_{r}m_{r}$ iterations, where $q_{r}$ is the number of rotations
performed in $l_{r}/m_{r}$. We then say that the orbit is of quasiregularity 
{\em type} $\tau =\dots ,(l_{r}/m_{r})_{q_{r}},(l_{r+1}/m_{r+1})_{q_{r+1}}\dots$ 
\cite{d2,d3}. As an example, Fig. 1 shows five orbit points, labeled by $
t=1,\dots ,5$, in two consecutive quasiregular segments $(0/1)_{3},(1/2)_{1}$.
\newline

In the case that the orbit is a PO, i.e., it satisfies (\ref{am}), its type
can be written in a more compact form \cite{bd,d2}. Clearly, a PO can visit
only a finite number ($d$)\ of resonances on the torus $0\leq x,\ p<2\pi $
[by taking also $p_{t}$ modulo $2\pi $ in (\ref{map})]. Thus, on the
cylinder, it will generally visit a set of $d$ resonances $\left\{
l_{r}/m_{r}\right\} _{r=1}^{d}$ and all the translates $\left\{
l_{r}/m_{r}+bw_{\tau }\right\} _{r=1}^{d}$ of this set in the $p$-direction,
where $b$ takes all the integer values and $w_{\tau }$ is some integer
related to $w$, see below. The type $\tau $ of the PO must be then
essentially the repetition of a ``block'' $\Gamma $, $\tau =\dots ,\Gamma
(-w_{\tau }),\Gamma (0),\Gamma (w_{\tau }),\Gamma (2w_{\tau }),\dots $,
where $\Gamma (bw_{\tau })=(l_{1}/m_{1}+bw_{\tau })_{q_{1}},\dots ,
(l_{d}/m_{d}+bw_{\tau })_{q_{d}}$. If the periodic cycle is completed 
exactly after visiting one block, one has $w=w_{\tau }$ and the period $
n=n_{\tau }=\sum_{r=1}^{d}q_{r}m_{r}$. Generally, however, the periodic
cycle is completed only after visiting more than one block, say $c$ blocks.
Then, $w=cw_{\tau }$ and $n=cn_{\tau }$. The type $\tau $ of the PO will be
thus specified by $(\Gamma ;w_{\tau },c)$, where $\Gamma $ stands, e.g., for 
$\Gamma (0)$. The average acceleration per iteration is $2\pi w/n=2\pi
w_{\tau }/n_{\tau }$.\newline

\begin{center}
{\bf III. ROTATING ACCELERATOR-MODE ISLANDS (RAIs)}\\[0pt]
\end{center}

If a period-$n$ PO is stable, each of its $n$ points is the ``center'' of an
island in a chain of $n$ islands. As shown in Ref. \cite{bd}, an island must
lie entirely within the zone of some resonance $l/m$. If this zone is the
principal one, ${\cal Z}^{(0)}(l/m)$, the island will be either outside the
turnstiles or completely within the turnstile overlap (TO) of $l/m$ with
another resonance $l^{\prime }/m^{\prime }$ \cite{bd}. Thus, the island will
always lie entirely in the basic region [resonance zone (outside the
turnstiles) or TO] where its center lies, so that one can characterize the
island chain by the type $(\Gamma ;w_{\tau },c)$ of its central PO. For
example, the well-known period-1 AM islands arising for\ $K>$ $2\pi |w|$ 
\cite{bvc,krw,ikh} must all lie within first-order ($m=1$) resonances and
their type is $\tau =((0/1)_{1};w,1)$; this is because $n=cn_{\tau
}=c\sum_{r=1}^{d}q_{r}m_{r}$ implies, for $n=1$, that $c=d=q_{1}=m_{1}=1$.
Since the fraction of phase space occupied by the $m=1$ resonances
approaches $100\%$ as $K$ increases \cite{mmp}, and is already significant
for $K>K_{{\rm c}}$, it is natural to ask about the existence of RAIs, i.e.,
AM islands visiting resonances of order $m>1$.\newline

To answer this question, we have carefully examined all the data on
standard-map AM islands of which we are aware. Most of this data appears in
Refs. \cite{krw,ikh,shi}; apparently new AM islands have been also
considered. The type of an island chain or of its central PO was accurately
determined by using the efficient method introduced in Ref. \cite{d3}.
Briefly, this method is based on the fact, proven in Ref. \cite{d3}, that
one can always find a sawtooth map $M_{{\rm s}}$ [i.e., the map (\ref{map})
with $\sin (x)$ replaced by a sawtooth function and with $K$ replaced by a
properly chosen parameter $K_{{\rm s}}$] such that for each orbit ${\cal O}$
of $M$ there exists an orbit ${\cal O}_{{\rm s}}$ of $M_{{\rm s}}$ visiting
the {\em same} resonances as those visited by ${\cal O}$. Since the
boundaries of the resonances of $M_{{\rm s}}$ are given by simple analytic
expressions \cite{saw}, this allows one to determine the type of ${\cal O}$
without calculating the complicated resonance boundaries of $M$.\newline

Our results are presented in Tables I and II. In Table I, we give the type $
\tau $ of many AM island chains for $K_{{\rm c}}<K<2\pi $. These chains,
most of which appear in Table I in Ref. \cite{ikh}, were chosen in a
well-defined and natural way, i.e., they have at least one island lying on $
p=0$ (see also next section), except for chains no. 7, 8 which are given for
future reference; initial conditions $(x_{0},p_{0})$ within the islands are
specified, as well as the values of $K$, $n$, and $w$. We see that more than
half of these island chains (no. 1-12, 15, 16), mostly at the smaller values
of $K$, are RAIs visiting resonances of order $m=2$ and/or $m=3$. The
central PO for RAI no. 6, with $n=5$ and $w=1$, consists of the five points
shown in Fig. 1. As another example, we show in Fig. 2 the central PO for
RAI no. 1, with $n=11$ and $w=1$, together with the four resonances visited, 
$l/m=0/1,1/3,1/2,2/3$. RAI no. 5, with $n=10$, emerges by period-doubling
bifurcation from RAI no. 4 with $n=5$. This is reflected in the fact that
the types of RAIs no. 4 and 5 have the same basic block $\Gamma $ but $c=w=2$
for RAI no. 5, in contrast with $c=w=1$ for RAI no. 4. Similarly, RAI no. 8,
with $n=8$, emerges by period-doubling bifurcation from RAI no. 7 with $n=4$.
\newline

In Table II, we give the type $\tau $\ of AM\ islands with $n=1,2,3$ for $
2\pi <K<20$. Most of these islands, those with $n=1,2$, were selected from
Table I in Ref. \cite{krw}. As in that work, we give in Table II the
stability interval $(K_{1},K_{2})$ of each island and corresponding initial
conditions $(x_{0},p_{0})$ for $K=K_{1,2}$; the values of $n$ and $w$ re
also shown. Islands no. 2, 6, and 9 emerge by period-doubling bifurcation
from islands no. 1, 5, and 8, respectively. All these islands lie in $m=1$
resonances. The only AM islands that we were able to identify as RAIs in
this $K$ interval are no. 3, 4, and 7. These RAIs are ``pure'', i.e., they
visit resonances of the same order $m=2$ (RAIs no. 3 and 7) or $m=3$ (RAI
no. 4). The latter RAI is apparently a new AM island. Figs. 3-5 show the
central PO for each of these pure RAIs, together with the resonances visited
and the turnstiles responsible for the transition from resonance $1/m$ to
resonance $1/m+w$; this transition causes the acceleration.\newline

RAIs can be very small islands. As examples, we show in Fig. 6 one island of
each of the RAI chains to which we refer in Figs. 2-5.\newline

\begin{center}
{\bf IV. CHAOTIC TRANSPORT IN THE PRESENCE OF RAIs}\\[0pt]
\end{center}

We now briefly study the effect of RAIs on chaotic transport. Given an
ensemble ${\cal E}$ of initial conditions $(x_{0},p_{0})$ in phase space for 
$K>K_{{\rm c}}$, the transport of ${\cal E}$ is usually measured by the time
evolution of $\left\langle (p_{t}-p_{0})^{2}\right\rangle _{{\cal E}}$,
where $\left\langle \ \right\rangle _{{\cal E}}$ denotes average over ${\cal
E}$. In the absence of AM islands, with ${\cal E}$ lying entirely within the
connected chaotic region, $\left\langle (p_{t}-p_{0})^{2}\right\rangle _{
{\cal E}}\approx 2Dt$ for large $t$, where $D$ is the chaotic-diffusion
coefficient \cite{bvc}. In Ref. \cite{ikh}, ${\cal E}$ was naturally chosen
as a physical ensemble of well-defined angular momentum $p=p_{0}$, $0\leq
x_{0}<2\pi $, and the quantity $D_{{\cal E},t}(K)=\left\langle
(p_{t}-p_{0})^{2}\right\rangle _{{\cal E}}/(2t)$ was calculated at fixed
large $t$ as a function of $K$, for $K_{{\rm c}}<K<2\pi $. Whenever an AM
island crosses the line $p=p_{0}$ as $K$ is varied, $D_{{\cal E},t}(K)$
exhibits ``ballistic'' peaks, see Fig. 7 for $p_{0}=0$ and Ref. \cite{ikh}.
The AM islands on $p=0$ in Table I were determined in this way. Some of the
most significant peaks in Fig. 7 are due to RAIs, e.g., RAIs no. 6-8, 10,
15, 16. In particular, the peak due to RAI no. 16 is relatively broad in $K$.
\newline

If ${\cal E}$ is an ensemble lying entirely within the connected chaotic
region and there exist AM islands exhibiting sufficient stickiness, one
observes an anomalous, superdiffusive chaotic transport, $\left\langle
(p_{t}-p_{0})^{2}\right\rangle _{{\cal E}}\propto t^{\mu }$ with anomalous
exponent $\mu $, $1<\mu <2$ \cite{ihkm,zen,bkwz,wbkz,ze,sd}. As far as we
are aware, this anomalous transport in the standard map was observed only
for AM islands whose central PO has period $n=1$ (such as islands no. 1, 5,
and 8 in Table II) or satellites of these islands. All these islands lie in $
m=1$ resonances and are not RAIs.\newline

To show that RAIs affect chaotic transport, we consider, as a first example,
the case of RAI no. 7 in Table I ($K=2.975$), see Fig. 8. The ensemble $
{\cal E}$ consists of the points $(x_{0},p_{0})$ with $p_{0}=0$ and $x_{0}$
taking $10^{5}$ values uniformly distributed in $[0,2\pi )$. This ensemble
is chaotic with the exception of $\sim 20\%$ of it lying in ordinary
(non-AM) islands within the $0/1$ resonance. The only source of anomalous
transport can be stickiness to the boundary of the RAI above, since no other
AM\ islands seem to exist for $K=2.975$. To verify that this RAI boundary is
indeed sticky, we have iterated the ensemble $t=10^{4}$ times and plotted
only the points $(x_{t^{\prime }},p_{t^{\prime }}$ mod$(2\pi ))$, for all $
t^{\prime }\leq t$ and with $p_{t}/(2\pi )>100$; since the RAI has central
period $n=4$, $p_{t}/(2\pi )$ cannot be larger than $t/4=2500$. The results
are shown in the inset of Fig. 8, and one can see a strong stickiness to the
RAI\ boundary. As a consequence, we observe a clearly superdiffusive chaotic
transport with anomalous exponent $\mu \approx 1.28$, see Fig. 9.\newline

As other examples, we have considered the much smaller RAIs shown in Figs.
6(b), 6(c), and 6(d). To verify the stickiness to the boundary of these
RAIs, we first chose ${\cal E}$ as a chaotic ensemble in the close
neighborhood of the RAI, see details in the captions of Figs. 6 and 10. The
time evolution of $\left\langle p_{t}-p_{0}\right\rangle _{{\cal E}}/(2\pi )$
was then calculated for sufficiently large $t$; the results are shown in
Fig. 10. We see that in a significant time interval (e.g., $t\leq 100$ in
Fig. 10(b)), $\left\langle p_{t}-p_{0}\right\rangle _{{\cal E}}/(2\pi )$
evolves essentially as if ${\cal E}$ were concentrated inside the RAI, i.e.,
it exhibits the steplike structure due to rotation in $m=2$ resonances (see
inset of Figs. 10(a) and 10(c)) or in $m=3$ resonances (see inset of Figs.
10(b)) and its initial average slope is $\sim w/n$. This is clear evidence
for stickiness to the RAI boundary, leading to chaotic flights. In the
course of time, more and more points of the ensemble leave the RAI boundary
and enter the chaotic region. Then, $\left\langle p_{t}-p_{0}\right\rangle _{
{\cal E}}/(2\pi )$ starts to saturate around some constant value which
should correspond to the center of a Gaussian distribution describing normal
chaotic diffusion. The saturation value of $\left\langle
p_{t}-p_{0}\right\rangle _{{\cal E}}/(2\pi )$ in Fig. 10(c) is much larger
than that in Figs. 10(a) and 10(b) due to the relatively large value of $
w/n=1$ and to the much stronger stickiness, as one can see by comparing Fig.
6(d) with Figs. 6(b) and 6(c).\newline

\begin{center}
{\bf V. DISCUSSION AND CONCLUSIONS}\\[0pt]
\end{center}

In this paper, we have established the existence of a most interesting
kind of stability island in the standard map, the RAI, exhibiting two
diametrically opposite dynamical behaviors: The rotational motion,
characteristic of the integrable ($K=0$) case, and the acceleration which
emerges only in the global-chaos regime of $K>K_{{\rm c}}\approx 0.9716$. As
indicated by Table I, RAIs appear to be abundant for sufficiently small $
K>K_{{\rm c}}$ but they also exist in strong-chaos regimes (Table II), where 
$m>1$ resonances are quite small and essentially all phase space is occupied
by $m=1$ resonances. For large $K>2\pi $, it is possible to have turnstile
overlap between resonances of the same order, e.g., resonances $1/m$ and $
1/m+w$, and pure RAIs can then arise, see Figs. 3-5. If the map (\ref{map})
is restricted to the torus $T^{2}:0\leq x,\ p<2\pi $, by taking also $p_{t}$
modulo $2\pi $, pure RAIs look precisely like rotational-resonance islands
in a near-integrable regime ($K\ll 1$). Stickiness to the boundary of RAIs,
especially of pure RAIs, leads to chaotic flights featuring a quasiregular
steplike structure due to the rotational motion within resonances, see Fig.
10. Such a quasiregular structure was observed recently \cite{d1} in the
weak-chaos regime of a perturbed pseudochaotic map. We have shown here that
it also occurs in strong-chaos regimes. It would be most interesting if one
could establish the existence of pure RAIs of arbitrarily large order $m$.
Such RAIs may give rise to chaotic flights with arbitrarily long
quasiregular steps; these flights were shown to be possible for the system
studied in Ref. \cite{d1}.\newline

There has been much interest recently \cite{fgr,qam,qam1} in the forced
standard map 
\begin{equation}
M_{\Omega }:\ \ p_{t+1}=p_{t}+K\sin (x_{t})+2\pi \Omega ,\ \ \ \
x_{t+1}=x_{t}+p_{t+1}\ \text{mod}(2\pi ),  \label{mapf}
\end{equation}
where $\Omega $ is a constant ``force''. It was shown \cite{fgr} that $
M_{\Omega }$ provides an approximate description of the vicinity of quantum
resonance for a periodically kicked particle in the presence of gravity. The
quantum resonant behavior of this system manifests itself in the so-called 
quantum AMs \cite{ao} which correspond to classical AM islands of $M_{\Omega }$. 
The map (\ref{mapf}) arises and was studied also in other contexts \cite{mapj,mapj1}. 
It is instructive to compare some AM islands of (\ref{mapf}) with the RAIs
in the standard map ($\Omega =0$). Consider, for simplicity, the case of
integer $\Omega \neq 0$ (the arguments and conclusions below can be easily
extended to the case of general rational $\Omega $). In this case, $
M_{\Omega }$ and the standard map obviously coincide if both maps are
restricted to the basic torus $T^{2}$. Thus, for sufficiently small $K$,
there exist rotational-resonance islands in any resonance $l/m$ of $
M_{\Omega }$ on $T^{2}$. On the cylinder, such an island will correspond to
an AM island: In one iteration of (\ref{mapf}), the island in zone ${\cal Z}
^{(t)}(l/m)$, $t=0,...,m-1$, will accelerate by jumping to zone ${\cal Z}
^{(t+1)}(l/m+\Omega )$ of resonance $l/m+\Omega $. Clearly, this behavior is
basically different from that of, say a pure RAI, which {\em remains}
(rotates) in resonance $l/m$ for $m$ (or a multiple of $m$) iterations
before jumping to resonance $l/m+w$. The rotational motion of RAIs, combined
with acceleration, is {\em not} featured by the AM islands above of $
M_{\Omega }$. It would be therefore interesting to study the fingerprints of
RAIs in the quantized standard map.\newpage

\begin{center}
{\bf ACKNOWLEDGMENTS}\\[0pt]
\end{center}

This work was partially supported by the Israel Science Foundation (Grant
No. 118/05).

Table I. Quasiregularity type $\tau =(\Gamma ;w_{\tau },c)$ of AM island
chains for $K_{{\rm c}}<K<2\pi $. The period $n$, jumping index $w$, and
initial conditions $(x_{0},p_{0})$ are also shown. 
\[
\begin{tabular}{|r|l|r|r|c|c|}
\hline
& $\ \ \ \ K$ & $\tau =(\Gamma ;\ w_{\tau },\ c)$ & $n\ $ & $w$ & $(x_{0},\
p_{0})/(2\pi )$ \\ \hline
$1$ & $2.1834$ & $((0/1)_{3},(1/3)_{1},(1/2)_{1},(2/3)_{1};\ 1,\ 1)$ & $11$
& $1$ & $(0.097,\ 0.0)$ \\ 
$2$ & $2.374$ & $((0/1)_{3},(1/2)_{1};\ 1,\ 1)$ & $7$ & $1$ & $(0.1085,\
0.0) $ \\ 
$3$ & $2.55097$ & $((0/1)_{4},(1/2)_{1},(1/1)_{3},(3/2)_{1};\ 2,\ 1)$ & $11$
& $2$ & $(0.102632,\ 0.0)$ \\ 
$4$ & $2.5875$ & $((0/1)_{3},(1/2)_{1};\ 1,\ 1)$ & $5$ & $1$ & $(0.098,\
0.0) $ \\ 
$5$ & $2.58867$ & $((0/1)_{3},(1/2)_{1};\ 1,\ 2)$ & $10$ & $2$ & $(0.099,\
0.00151)\ $ \\ 
$6$ & $2.63894$ & $((0/1)_{3},(1/2)_{1};\ 1,\ 1)$ & $5$ & $1$ & $(0.105,\
0.0)$ \\ 
$7$ & $2.975$ & $((0/1)_{2},(1/2)_{1};\ 1,\ 1)$ & $4$ & $1$ & $(0.33445,\
0.4)\ $ \\ 
$8$ & $2.9845$ & $((0/1)_{2},(1/2)_{1};\ 1,\ 2)$ & $8$ & $2$ & $(0.344,\
0.412)$ \\ 
$9$ & $3.34579$ & $((0/1)_{9},(1/2)_{1};\ 1,\ 1)$ & $11$ & $1$ & $(0.3851,\
0.0)$ \\ 
$10$ & $3.50287$ & $((0/1)_{5},(1/2)_{1};\ 1,\ 1)$ & $7$ & $1$ & $(0.64,\
0.0)$ \\ 
$11$ & $3.61283$ & $((0/1)_{9},(1/3)_{1},(1/1)_{2},(5/3)_{1};\ 2,\ 1)$ & $17$
& $2$ & $(0.336347,\ 0.0)$ \\ 
$12$ & $3.76991$ & $((0/1)_{5},(1/2)_{1};\ 1,\ 1)$ & $7$ & $1$ & $(0.6195,\
0.0)$ \\ 
$13$ & $3.78247$ & $((0/1)_{3},(1/1)_{2};\ 2,\ 1)$ & $5$ & $2$ & $(0.0667,\
0.0)$ \\ 
$14$ & $3.80761$ & $((0/1)_{3},(1/1)_{2};\ 2,\ 1)$ & $5$ & $2$ & $(0.07,\
0.0)$ \\ 
$15$ & $4.141$ & $((0/1)_{1},(1/2)_{1};\ 1,\ 1)$ & $3$ & $1$ & $(0.285,\
0.0) $ \\ 
$16$ & $4.66526$ & $((0/1)_{1},(1/2)_{1};\ 1,\ 1)$ & $3$ & $1$ & $(0.34805,\
0.0)$ \\ 
$17$ & $5.02654$ & $((0/1)_{3},(1/1)_{1},(2/1)_{2},(3/1)_{1};\ 4,\ 1)$ & $7$
& $4$ & $(0.04432,\ 0.0)$ \\ 
$18$ & $5.12079$ & $((0/1)_{2},(1/1)_{1};\ 2,\ 1)$ & $3$ & $2$ & $(0.277,\
0.0)$ \\ 
$19$ & $5.32499$ & $((0/1)_{3},(1/1)_{2};\ 2,\ 1)$ & $4$ & $2$ & $(0.055,\
0.0)$ \\ 
$20$ & $5.41924$ & $((0/1)_{2},(1/1)_{2},(2/1)_{1};\ 3,\ 1)$ & $5$ & $3$ & $
(0.338162,\ 0.0)$ \\ 
$21$ & $5.45066$ & $((0/1)_{2},(1/1)_{1};\ 2,\ 2)$ & $6$ & $4$ & $(0.22265,\
0.0)$ \\ 
$22$ & $5.51663$ & $((0/1)_{2},(1/1)_{1};\ 2,\ 1)$ & $3$ & $2$ & $(0.334,\
0.0)$ \\ 
$23$ & $5.79623$ & $((0/1)_{6},(1/1)_{1};\ 2,\ 1)$ & $7$ & $2$ & $(0.3255,\
0.0)$ \\ 
$24$ & $6.173229$ & $((0/1)_{3},(1/1)_{1},(2/1)_{1},(3/1)_{1};\ 4,\ 1)$ & $6$
& $4$ & $(0.0286,\ 0.0)$ \\ \hline
\end{tabular}
\]

\newpage Table II. Quasiregularity type $\tau =(\Gamma ;w_{\tau },c)$ of AM
island chains for $2\pi <K<20$. Also shown are the period $n$, the jumping
index $w$, and the stability interval $(K_{1},K_{2})$ of the island chain
with corresponding initial conditions $(x_{0},p_{0})$ for $K=K_{1,2}$.

\[
\begin{tabular}{|c|r|c|c|l|c|l|c|}
\hline
& $\tau =(\Gamma ;\ w_{\tau },\ c)$ & $n$ & $w$ & $\ \ \ \ K_{1}$ & $
(x_{0},\ p_{0})/(2\pi ),\ K=K_{1}$ & $\ \ \ \ K_{2}$ & $(x_{0},\
p_{0})/(2\pi ),\ K=K_{2}$ \\ \hline
$1$ & $((0/1)_{1};\ 1,\ 1)$ & $1$ & $1$ & $6.28319$ & $(0.2500,\ 0.0)$ & $
7.44840$ & $(0.34059,\ 0.0)$ \\ 
$2$ & $((0/1)_{1};\ 1,\ 2)$ & $2$ & $2$ & $7.44840$ & $(0.34059,\ 0.0)$ & $
7.71340$ & $(0.3845,\ 0.0928)$ \\ 
$3$ & $((1/2)_{1};\ 1,\ 1)$ & $2$ & $1$ & $8.67893$ & $(0.27999,\ 0.321577)$ & 
$8.68826$ & $(0.28696,\ 0.32717)$ \\ 
$4$ & $((1/3)_{1};\ 1,\ 1)$ & $3$ & $1$ & $8.91596$ & $(0.934594,\ 0.6522)$
& $8.91603$ & $(0.934641,\ 0.651819)$ \\ 
$5$ & $((0/1)_{1};\ 2,\ 1)$ & $1$ & $2$ & $12.56638$ & $(0.2500,\ 0.0)$ & $
13.1876$ & $(0.2990,\ 0.0)$ \\ 
$6$ & $((0/1)_{1};\ 2,\ 2)$ & $2$ & $4$ & $13.1876$ & $(0.2990,\ 0.0)$ & $
13.33848$ & $\ (0.3233,\ 0.0494)$ \\ 
$7$ & $((1/2)_{1};\ 2,\ 1)$ & $2$ & $2$ & $15.2394$ & $(0.268606,\ 0.295564)$ & $
15.24126$ & $(0.270938,\ 0.29762)$ \\ 
$8$ & $((0/1)_{1};\ 3,\ 1)$ & $1$ & $3$ & $18.84956$ & $(0.2500,\ 0.0)$ & $
19.26929$ & $(0.2832,\ 0.0)$ \\ 
$9$ & $((0/1)_{1};\ 3,\ 2)$ & $2$ & $6$ & $19.26929$ & $(0.2832,\ 0.0)$ & $
19.3728$ & $(0.2998,\ 0.0333)$ \\ \hline
\end{tabular}
\]

\newpage

\begin{center}
{\bf FIGURE CAPTIONS}\\[0pt]
\end{center}

Fig. 1. Solid lines: Resonances $0/1$ (region $LERF$) and $1/2$ (with principal
zone $L'E'R'F'$) for $K=2.63894$. The dashed-line segments $FBH$ and  $GAE$ form, 
together with the corresponding solid-line segments, the lower and upper turnstile, 
respectively, of $0/1$; similarly for the turnstiles of $1/2$. The shaded regions are the 
overlaps of the upper turnstile of $0/1$ with the lower turnstile of $1/2$. Also shown are
five orbit points labeled by the time index $t=1,...,5$. Since point 3 lies in $0/1$ within 
a turnstile overlap, its iterate (point 4) lies in zone 1 of $1/2$.\newline

Fig. 2. Regions bounded by solid lines: Resonances $0/1$, $1/3$, $1/2$, and 
$2/3$ (in ascending order) for $K=2.1834$. The dashed lines define the resonance
turnstiles. The points (filled circles) give the central PO of the AM island chain no. 1 in 
Table I; this is a RAI chain. The arrow indicates a point to which we shall refer in the 
caption of Fig. 6.\newline

Fig. 3. Regions bounded by solid lines: Resonances $1/2$ and $3/2$ for 
$K=8.68$. The dashed lines define the resonance turnstiles. The four points, labeled by
the time index $t=1,...,4$, give the central PO and its iterate for the AM island chain no. 
3 in Table II; this is a pure RAI chain.\newline

Fig. 4. Regions bounded by solid lines: Resonances $1/3$ and $4/3$ for 
$K=8.916$. The dashed lines define the resonance turnstiles. The six points, labeled by
the time index $t=1,...,6$, give the central PO and its iterate for the AM island chain no. 
4 in Table II; this is a pure RAI chain.\newline

Fig. 5. Regions bounded by solid lines: Resonances $1/2$, $3/2$, and $5/2$ 
for $K=15.24035$. The dashed lines define the resonance turnstiles. The four points, 
labeled by the time index $t=1,...,4$, give the central PO and its iterate for the AM island 
chain no. 7 in Table II; this is a pure RAI chain. Point 2 in $1/2$ is mapped into point 3 
in $5/2$ by turnstile overlap, ``jumping over" $3/2$; this leads to $w=2$.\newpage

Fig. 6. (a) RAI surrounding the point indicated by an arrow in Fig. 2 
($K=2.1834$); the region $(x_1,x_2)/(2\pi )\times (p_1,p_2)/(2\pi )$ covered by the 
graph is $(0.09694, 0.097165)\times (-5.5\cdot 10^{-5},5.5\cdot 10^{-5})$. (b) RAI 
surrounding point 1 in Fig. 3 ($K=8.68$), graph region is $(0.9582,0.9592)\times 
(0.674,0.68)$. (c) RAI surrounding point 2 in Fig. 4 ($K=8.916$), graph region is 
$(0.019943,0.0199537)\times (0.085314,0.085334)$. (d) RAI surrounding point 1 in 
Fig. 5 ($K=15.24035$), graph region is $(0.97312,0.97326)\times (0.7019,0.704)$.\newline

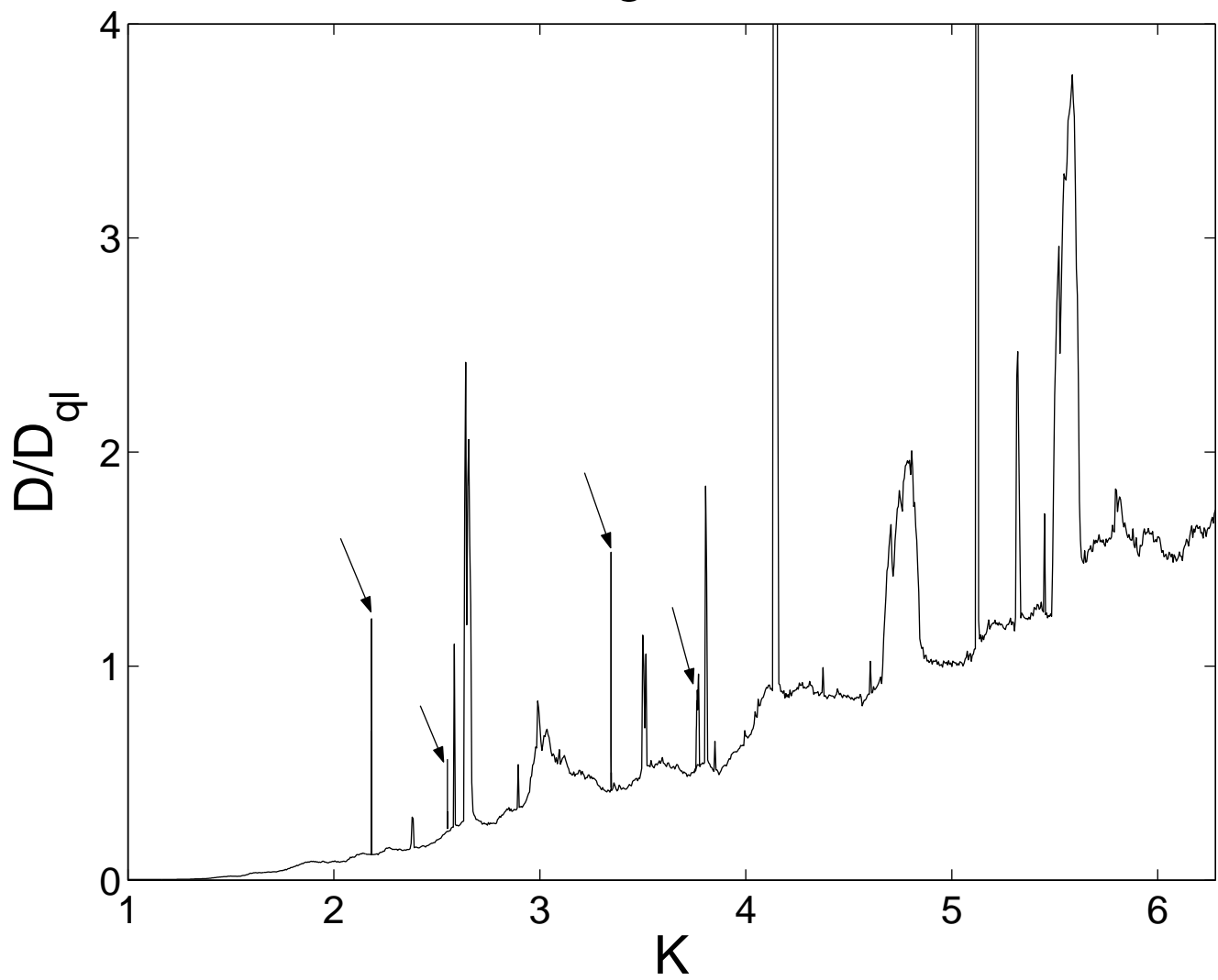
Fig. 7. Plot of $D_{{\cal E},t}(K)/D_{\rm ql}$ for $t=500$ and $K\in (1,2\pi )$ 
($\Delta K=5\cdot 10^{-3}$), where $D_{\rm ql}=K^2/4$ is the quasilinear diffusion 
coefficient. The ensemble ${\cal E}$ consists of the points $(x_{0},p_{0})$ with 
$p_{0}=0$ and $x_{0}$ taking $20000$ values uniformly distributed in $[0,2\pi )$. The 
peaks indicated by arrows correspond to the RAIs no. 1, 3, 9, and 12 in Table I and 
were produced using $t=20000$ iterations and a finer grid in both $K$ and $x_0$.\newline

Fig. 8. Solid lines: Resonances 0/1 and 1/2 for $K=2.975$. Shown is the RAI 
chain no. 7 in Table I, labeled by the time index $t=1,...,4$. A magnification of island 
no. 1 appears in the inset, showing strong stickiness of chaotic orbits to the RAI 
boundary.\newline

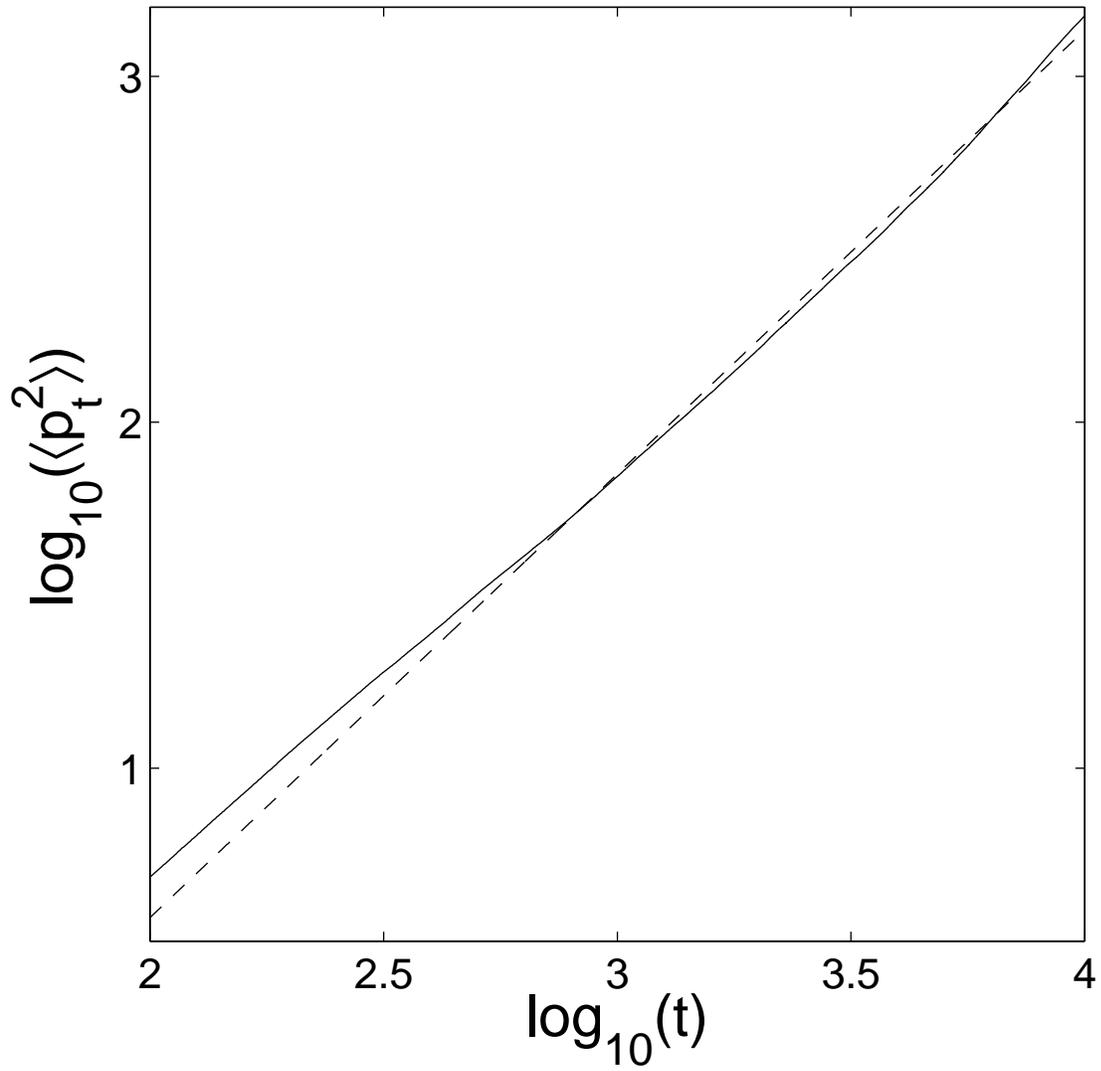
Fig. 9. Solid line: Log-log plot of $\langle p_t^2\rangle _{\cal E}$ for $K=2.975$ 
(see more details in the text). Dashed line: Linear fit to the solid line, with slope 
$\mu\approx 1.28$.\newline

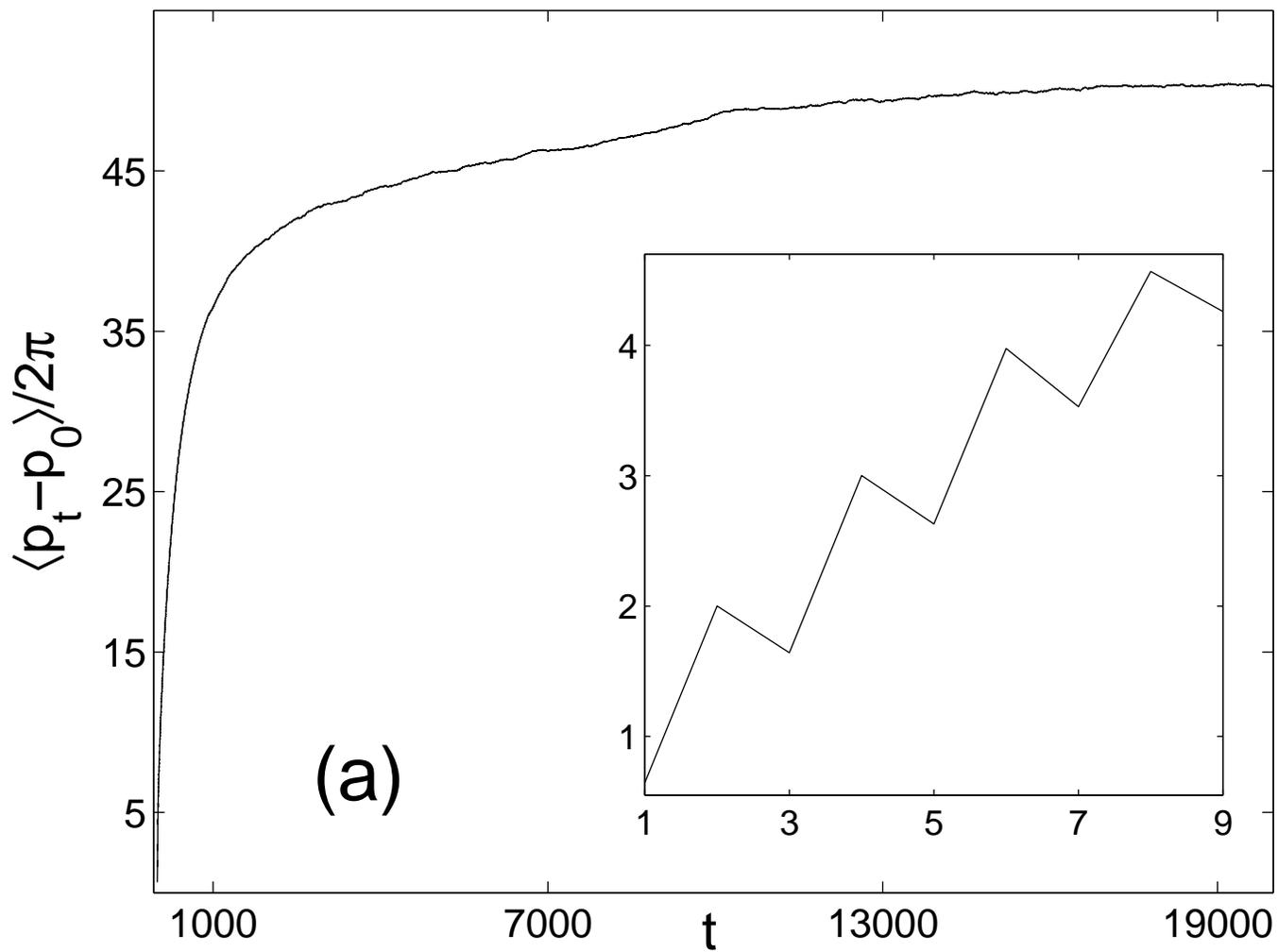
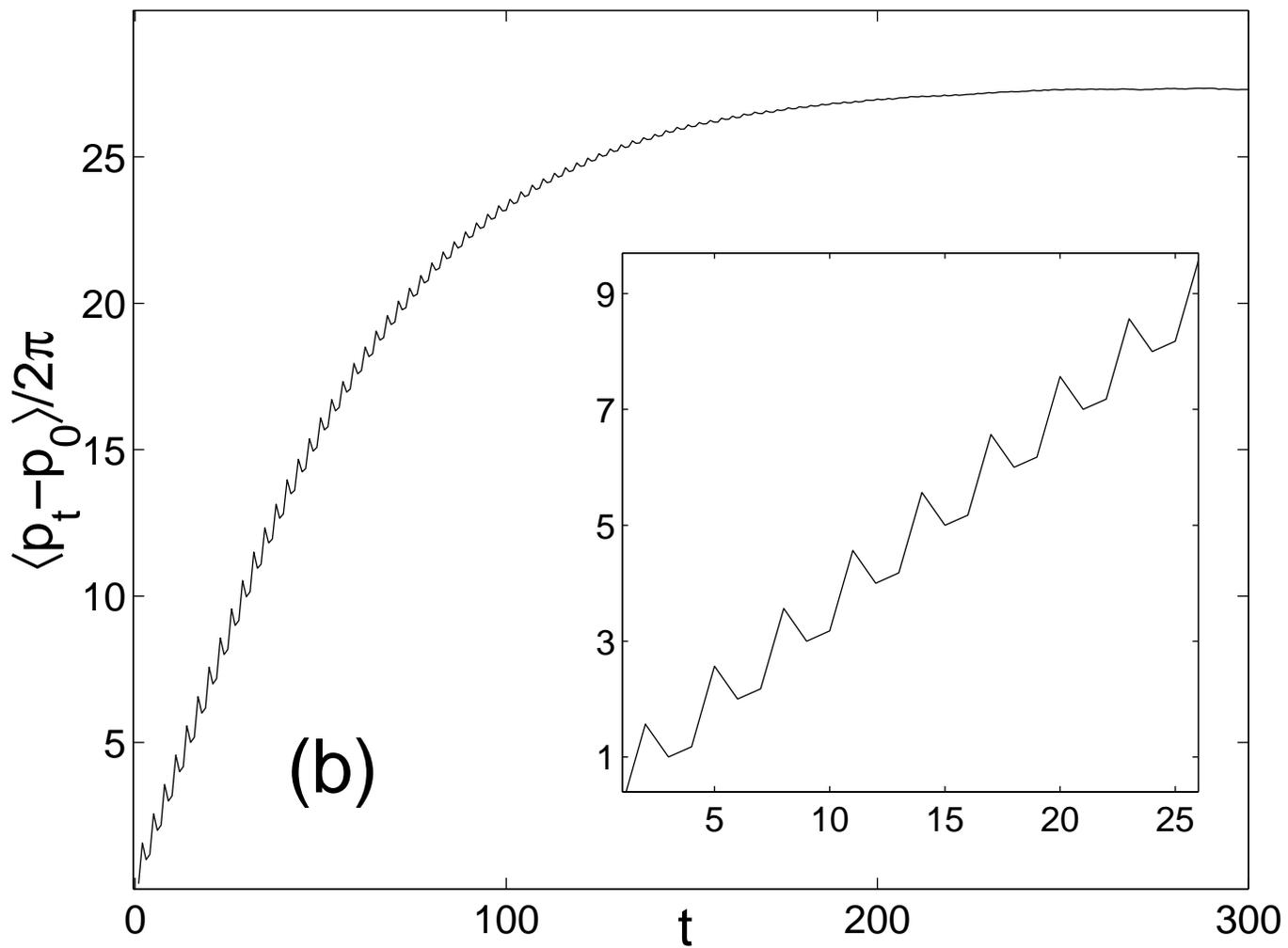
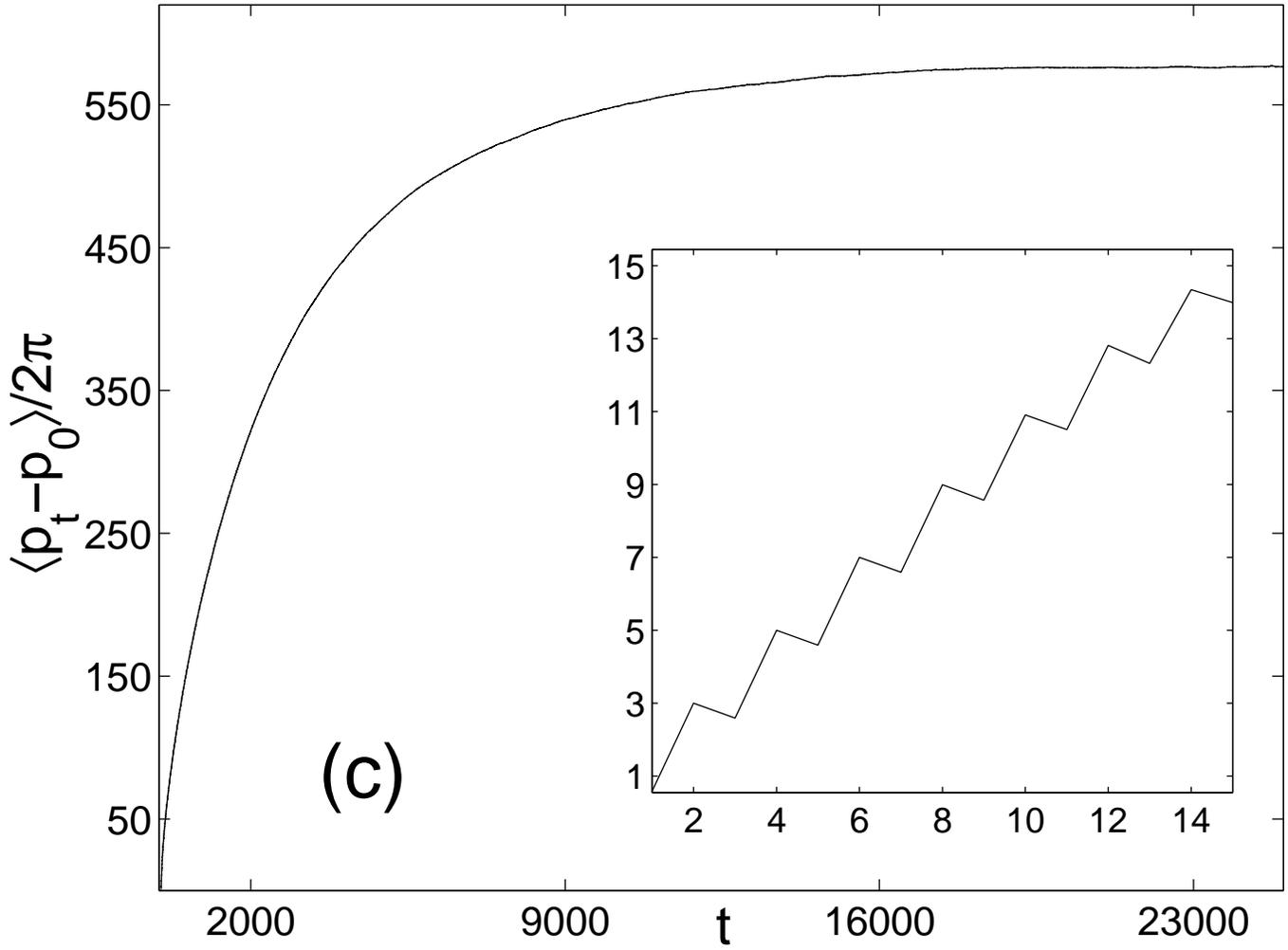
Fig. 10. Time evolution of $\langle p_t-p_0 \rangle _{\cal E}/(2\pi )$, where 
${\cal E}$ is an ensemble of chaotic initial conditions (ICs) $(x_0,p_0)$ extracted from a 
grid covering the graph region specified in the caption of: (a) Fig. 6(b) (19276 chaotic 
ICs out of 26867), $K=8.68$; (b) Fig. 6(c) (15639 chaotic ICs out of 21600), $K=8.916$; 
(c) Fig. 6(d) (10693 chaotic ICs out of 11788), $K=15.24035$. The inset in (a), (b), and 
(c) shows a magnification of the first iterates with average slope $\sim w/n=1/2,\ 1/3$, 
and $1$, respectively. The steplike structure due to the stickiness to the boundaries of 
the pure RAIs is quite evident in all cases.\newline

\end{document}